\documentclass[a4paper,10pt]{article}
\usepackage{graphicx}
\usepackage[labelfont={it,bf},font=it]{caption}
\usepackage{amsmath}
\usepackage{amsfonts}

\usepackage{pslatex}

\makeatletter
\renewcommand\section{\@startsection {section}{1}{\z@}%
    {-3.5ex \@plus -1ex \@minus -.2ex}%
    {2.3ex \@plus.2ex}%
    {\normalfont\bfseries\MakeUppercase}}
\makeatother

\begin{document}

\pagestyle{empty}

\begin{center}
{\fontsize{16}{16pt} \bf
 Coherent beam shaping using two-dimensional photonic crystals}
\vspace{12pt}\vspace{12pt}\\
{\bf Denis Gagnon, Joey Dumont and Louis J. Dub\'e*}\\
{\it D\'epartement de Physique, de G\'enie Physique, et d'Optique, Facult\'e des Sciences et de G\'enie \\
Universit\'e Laval, Qu\'ebec, QC G1V 0A6, Canada\\
*Corresponding author: ljd@phy.ulaval.ca}
\end{center}
\vspace{-0.6cm} \vspace{6pt}

\section*{Abstract}

Optical devices based on photonic crystals such as waveguides, lenses and beam-shapers, have received considerable theoretical and experimental attention in recent years. The production of these devices has been facilitated by the wide availability of silicon-on-insulator fabrication techniques. In this theoretical work, we show the possibility to design a coherent PhC-based beam-shaper. The basic photonic geometry used is a 2D square lattice of air holes in a high-index dielectric core. We formulate the beam shaping problem in terms of objective functions related to the amplitude and phase profile of the generated beam. We then use a parallel tabu search algorithm to minimize the two objectives simultaneously. Our results show that optimization of several attributes in integrated photonics design is well within reach of current algorithms.

\section{Introduction}
Integrated photonics components such as waveguides \cite{Frandsen2004}, beam-splitters \cite{Pottier2006}, slow-light devices \cite{Baba2008}, lenses \cite{Marques-Hueso2013} and beam-shapers \cite{Gagnon2012} have received considerable attention in the last decade. These components can be manufactured using two-dimensional (2D) photonic crystals (PhCs). More specifically, PhCs based on air holes in a planar waveguide based dielectric core have been successfully manufactured in silicon-on-insulator material using microfabrication techniques \cite{Frandsen2004, Pottier2006, Baba2008}. Prior to fabrication, a design step often involves selecting a number of geometric parameters related to the PhC and performing optimization of a cost function over a large solution space. For instance, a basic photonic lattice can be defined and holes allowed to be present or absent, thereby enabling a binary encoding of the solution space. Optimization may then be carried out using a standard genetic algorithm \cite{Gagnon2012,
 Sanchis2004, 
Vukovic2010}.

This conference paper is concerned with the generation of arbitrary coherent beam profiles using engineered 2D PhCs. In other terms, we seek to transform a known input beam into another beam of controlled amplitude \textit{and} phase profile.
Those two conditions constitute a \textit{multiobjective optimization problem} (MOP), which must be solved by sampling the set of optimal solutions, commonly known as the Pareto set \cite{Talbi2009}. The first part of this contribution briefly describes the optimization procedure for a specific target, namely the generation of coherent Hermite-Gauss beam profiles using a 2D PhC slab. In the latter part, we describe the MOP solving procedure used and present some solutions offering an acceptable trade-off between the two aforementioned objectives (amplitude and phase profile of the beam).

\section{Coherent beam shaping problem}
Laser beam shaping is defined as redistributing the irradiance and phase of a beam \cite{Dickey2005}. This contribution is concerned with finding a PhC configuration which, when illuminated with a Gaussian beam, produces a scattered wavefunction that matches a desired profile in a given plane. The beam shaping problem can be formulated as the minimization of the following integral 
\begin{equation}
g_1 = \dfrac{\int \big||u(x_0,y)|^2 - |\bar{u}(x_0,y)|^2 \big| dy}{\int |\bar{u}(x_0,y)|^2 dy } 
\end{equation}
where $x_0$ is the location of the target plane, $u(x_0,y)$ is the computed EM field on the target plane, $\bar{u}(x_0,y)$ is the desired beam at the device output (the $x-$axis is the beam propagation axis). It was recently proposed to solve this optimization problem using a combination of multiple scattering computations and a genetic algorithm. Further details can be found in Ref. \cite{Gagnon2012}. However, the minimization of $g_1$ does not take into account the phase profile of the beam, only the amplitude, or irradiance distribution. This kind of optimization problem is called \textit{incoherent beam shaping}. As a result, optimized beams may exhibit large transverse phase fluctuations, which in turn results in an poor field depth. This is a major impediment to applications such as atom guiding \cite{Molina-Terriza2007} and microscopy \cite{Olivier2012}, where beams with large field depths (low divergence) are needed. In order to achieve \textit{coherent beam shaping}, we must define another objective 
function related to phase fluctuations of the transverse profile. Our proposal is to minimize the following integral
\begin{equation}
g_2 = \dfrac{\int \big| \mathrm{Im} [ u(x_0,y) e^{-i\phi(x_0,0)} ]  \big|^2 dy}{\int |\bar{u}(x_0,y)|^2 dy }
\end{equation}
where $\tan \phi(x,y) = \mathrm{Im}  [ u(x,y) ] / \mathrm{Re}  [ u(x,y) ]$. The value of $g_2$ is zero for a collimated beam (plane phase front), and increases with the number of oscillations in the phase front.

To illustrate the potential of multiobjective optimization techniques in integrated photonics design, we apply them to the problem of coherent beam shaping. The desired output beam and target plane can be arbitrary \cite{Gagnon2012}. For illustrative purposes, we choose to generate Hermite-Gauss beam profiles of half-width $w$ and order 2 at the device output, that is
\begin{equation}
\bar{u}(x_0,y) = \left[4 \xi^2 - 2 \right]^2 \exp \left( -\xi^2  \right)
\end{equation}
where $\xi = \sqrt{2} y / w$. We also seek a normalized beam profile, since backscattering losses are mostly unavoidable in PhC devices \cite{Gagnon2012}.
The basic scatterer geometry is a $13 \times 10$ square lattice of air holes embedded in a medium of index $n = 2.76$, for a total of $N_s = 130$ possible scatterers. The diameter of all holes is set to $D = 0.6 \Lambda$, where $\Lambda$ is the lattice constant. For definiteness, we prescribe our incident beam as a TM-polarized non-paraxial Gaussian beam with a half-width $w_0=2.5\Lambda$ and a wavenumber $k_0 = 1.76 / \Lambda$ for a Rayleigh distance $x_R =k_0 w_0^2 /2 = 5.48 \Lambda$. Moreover, a mirror symmetry across the $x$ axis is taken into account, resulting in $2^{70}$ possible solutions, or $\sim 10^{21}$.

\section{Multiobjective optimization results}
The simultaneous minimization of multiple objective functions (in this case $g_1$ and $g_2$) can be tackled using a set of techniques known as \textit{multiobjective optimization}. Those techniques have been used in economics for several decades, and more recently in sciences and engineering. Since it is not generally possible to minimize the entire set of objective functions, the solution of a MOP is not a single solution, but rather a set of solutions termed \textit{Pareto optimal} solutions. A solution is called Pareto optimal if it is not possible to improve a given objective without deteriorating at least another \cite{Talbi2009}. The resolution of a MOP therefore amounts to obtaining the Pareto optimal set of solutions. Since many real-world problem including photonics design are NP-hard, the sampling of the Pareto set must be achieved using metaheuristics such as the genetic algorithm \cite{Sanchis2004, Vukovic2010}. The genetic algorithm (GA) is a evolutionary algorithm based on a stochastic 
exploration 
of the 
solution space. In this work, we rather use the parallel tabu 
search (PTS) algorithm \cite{Crainic1997} to efficiently solve the optimization problem. 

The PTS algorithm is based on a group of search processes exploring the solution space in a parallel fashion \cite{Crainic1997}. One iteration of a tabu search process begins by generating all neighbours of the current solution. The best possible neighbour (best possible value of the objective function) is then chosen for the next iteration, unless that move is prohibited by the \textit{tabu list}. This list constitutes the short-term memory of the algorithm and prevents cyclic trajectories in the solution space \cite{Talbi2009}. Unlike the GA, the PTS algorithm is a deterministic search algorithm; only the initial solutions for each search process are randomly initialized. The tabu search also involves less adjustable parameters, while the parallel nature of the search ensures a covering of the solution space as good as that of the GA. To our knowledge, this is the first application of this sort of techniques to the field of integrated photonics.

In order to take our multiple objectives ($g_1$ and $g_2$) into account, we make use of the so-called \textit{aggregation method} to sample the Pareto optimal set \cite{Talbi2009}. In a nutshell, one defines an aggregated fitness function $f$ in the following fashion
\begin{equation}
f = (1 - \alpha) \frac{g_1}{g_1^{max}} + \alpha \frac{g_2}{g_2^{max}}
\end{equation}
where $g_1^{max}$ and $g_2^{max}$ are heuristic upper bounds of both objective functions. The Pareto front (location of the set of optimal solutions) is then sampled by running several PTS using different values of the parameter $\alpha$. This has the effect of increasing or decreasing the relative importance of $g_2$, thereby steering the search towards different regions of the Pareto optimal set. In the case of the beam shaping problem, this simple implementation of a MOP amounts to solving a single-objective minimization problem over a $70-$dimensional space, multiple times with different fitness functions. The sampling of the Pareto front was performed using 7 different values of $\alpha \in [0.0,0.425]$. For each of those values, 48 tabu search processes are executed in parallel, each totalling 5000 iterations. This set of search processes yields a number of final solutions, out of which we extract the optimal solutions. A solution is Pareto optimal if there is no solution found that is characterized by 
a lower value of both $g_1$ and $g_2$.
\begin{figure}
\centering
 \includegraphics[width=0.5\textwidth]{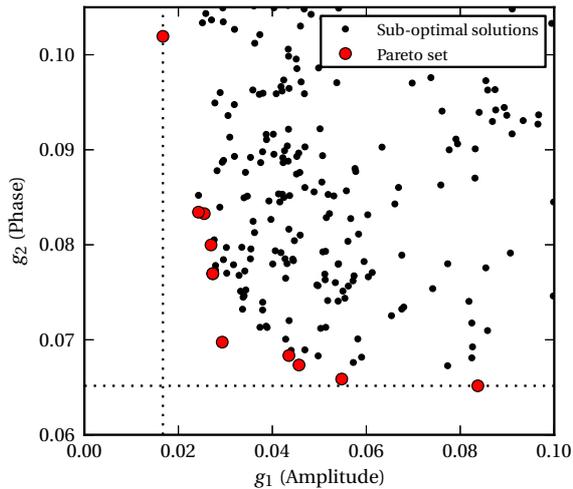}
\caption[Pareto front for the coherent beam shaping profile]{Pareto front for the coherent beam shaping problem. The sampling of the front is achieved via a series of PTS computations, using $\alpha \in [0.0,0.425]$. The dotted lines indicate the best possible value for each of the two objectives.}
\label{fig:pareto}
\end{figure}
\begin{figure}
\centering
 \includegraphics[]{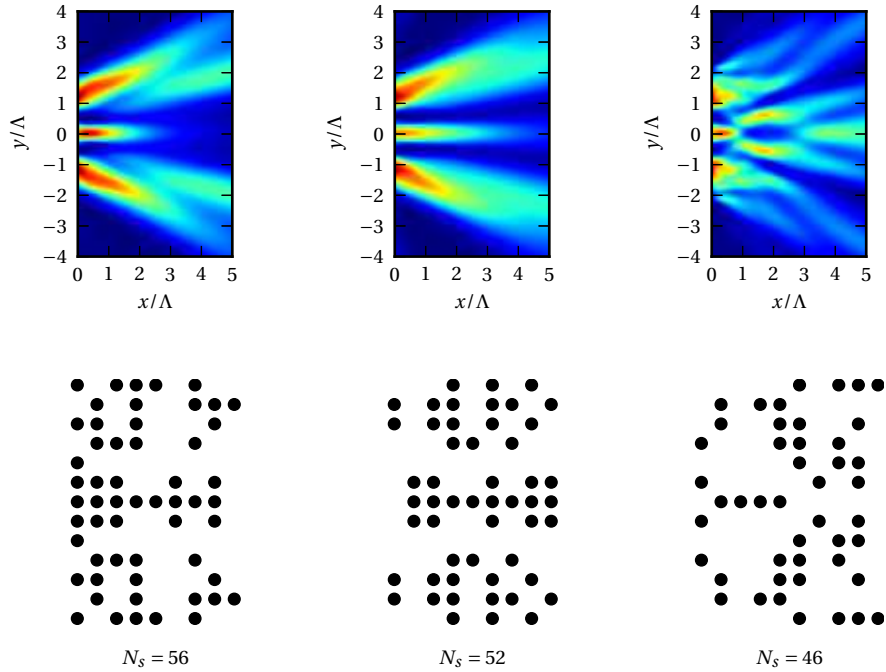}
\caption[Three candidate solutions for the coherent beam shaping problem]{Three candidate solutions for the coherent beam shaping problem. \textbf{Top row:} Generated beams. \textbf{Bottom row:} Associated PhC configurations with number of scatterers $N_s$. \textbf{Left column:} Solution with ($g_1,g_2$) = ($0.017,1.01$). \textbf{Center column:} Solution with ($g_1,g_2$) = ($0.026, 0.83$). \textbf{Right column:} Solution with ($g_1,g_2$) = ($0.043, 0.68$). }
\label{fig:beams}
\end{figure}

Pareto optimal results, as well as the various sub-optimal (non-Pareto) solutions, are presented on fig. \ref{fig:pareto}. The set of Pareto solutions found offer the best possible compromise between an accurate amplitude profile and a uniform phase front. As illustration, three sample Pareto solutions are presented on fig. \ref{fig:beams}. The configuration on the left offers the most accurate reproduction of a Hermite-Gauss beam profile. However, the non-uniformity of the phase front results in a poor field depth. Inversely, the configuration on the right exhibits a mostly uniform phase front, but is not close to the amplitude profile of a order 2 Hermite-Gauss beam. The central configuration exhibits the best trade-off between the two required attributes, keeping a Hermite-Gaussian profile over a greater distance.

It should be noted that the ``optimality'' of the solutions strongly depends on a given application. The results on fig. \ref{fig:pareto} rather give a snapshot of the potential of the pre-defined photonic lattice. We also stress the fact that multiobjective techniques can readily take other objectives into account. For instance, one could seek to minimize the backscattering losses associated to the finite photonic cystal. However, since at least all Pareto solutions of a $n-$ objective problem are necessary solutions of the same problem with a greater number of objectives, the computational cost increases accordingly \cite{Talbi2009}.

\section{Conclusion}

In this contribution, we have reported the possibility to control the coherent profile (amplitude and phase) of the output beam in a PhC-based integrated beam shaping device. The shaping problem was formulated in terms of two objective functions, one for the amplitude and one for the phase of the transformed beam. A parallel tabu search algorithm was used to sample the Pareto front (optimal solutions) of the resulting multiobjective problem. Our results show that multiobjective optimization of integrated photonics devices is within reach of currently available algorithms.

The authors acknowledge financial support from the Natural Sciences and Engineering Research Council of Canada (NSERC) and computational resources from Calcul Qu\'ebec.


\end{document}